\documentclass[prb,aps,twocolumn,dcolumn,superscriptaddress,showpacs,amsfonts,amsmath,amssymb,showkeys,floatfix]{revtex4}
\usepackage{bm}
\usepackage{graphicx}% Include figure files
\begin{document}
\title{Equilibrium spin-glass transition of magnetic dipoles with random anisotropy axes}
\date{\today}
\author{Julio F. Fern\'andez}
\affiliation{Instituto de Ciencia de Materiales de Arag\'on, CSIC and Universidad de Zaragoza, 50009-Zaragoza, Spain}
%\altaffiliation{IVIC}
\email[E-mail address: ] {jefe@Unizar.Es}

\date{\today}
%\keywords{spin reorientation, films, anisotropy, Monte Carlo}

\begin{abstract}
We study fully occupied lattice systems of classical magnetic dipoles which point along random axes. Only dipolar interactions are considered. From tempered Monte Carlo simulations, we obtain numerical evidence  that supports the following conclusions: in three dimensions, (a) there is an equilibrium spin glass phase at temperatures below $T_c$, where $k_BT_c= (0.86\pm 0.07)\varepsilon_d $ and $\varepsilon_d$ is a nearest neighbor dipole-dipole interaction energy, (b) in the spin glass phase the overlap parameter is approximately given by $\sqrt{1-T/T_c}$, and (c) the correlation length $\xi$ diverges at $T_c$ with a critical exponent $\nu =1.5\pm0.5$; in two dimensions $\xi$ diverges at or near $T=0$ and $\nu=3\pm1$.
\end{abstract}

\maketitle

%\pacs{75.45.+j, 75.50.Xx,75.70.-i }
%\keywords{spin reorientation, films, anisotropy, Monte Carlo}

\section{introduction}
Several decades after the experimental discovery of spin glasses,\cite{discovery} convincing numerical evidence for an equilibrium phase transition between the paramagnetic and spin-glass phases of the random bond Ising \cite{kawa,balle} model in three dimensions is at last available. Somewhat more controversial evidence is also available for the Heisenberg model.\cite{japon,apy} No such results that we know of exist for systems in which dipole-dipole interactions dominate. This is in spite of all the interest that has arisen in these systems since nanosized magnetic particles have become experimentally available. \cite{nanom} Randomness, one of the two essential ingredients for spin-glass behavior, can arise from spatial disorder,\cite{so} which in turn, most often, brings about random magnetic anisotropies. One might naively expect that the long range nature of dipolar interactions would only strengthen the spin glass phase that is observed in the random (nearest neighbor) bond Ising model. However, recent results from computer simulations suggest that an equilibrium spin glass phase does not obtain in a spatially disordered system of magnetic dipoles which point along parallel axes.\cite{clare} The reason for this somewhat unexpected result may be the nature of frustration that is peculiar to dipolar systems. 
In them, there is frustration whether they are spatially ordered or not. It is precisely because of this that ferro- or antiferro-magnetism prevails in well ordered crystalline dipolar systems depending delicately on lattice geometry.\cite{nos0}
On the other hand, spin-glass like behavior has been observed in experiments \cite{irrev,control,expymcaging} and in simulations \cite{eaging,mcaging,ulrich, labarta,bunde} of dipolar systems with random anisotropies, but all this evidence comes from {\it out of equilibrium} phenomena, as exhibited by time dependent susceptibilities, nonexponential relaxation, and aging.\cite{eaging}

We study the equilibrium behavior of systems of interacting magnetic dipoles which are oriented along random anisotropy axes in two and three dimensions (D).
 This random axes dipolar (RAD) model is like the old model of
Harris, Plischke, and Zuckerman, \cite{RAM} except that we deal with dipole-dipole, rather than nearest neighbor (nn) interactions. Some motivation for the RAD model comes from the fact that anisotropy energies in nanoparticle assemblies are often \cite{notalways} much larger than the dipole-dipole interaction energy between two nearest neighbors. As in an Ising model, spins in the RAD model can only point ``up'' or ``down'' along each one of their own axes, as is discussed in Refs. [\onlinecite{bunde,kirk,cieplak}].
Two independent random numbers per site are needed to determine all axes directions, which is the same number as for a nn random bond Ising model on a square lattice, though the interaction range is of course quite different.

When we simulate the time evolution of the RAD model, we flip each spin up and down along its own axis. We thus make no attempt to simulate how each individual spin overcomes large anisotropy barriers. Rather, we expect our simulations to mimic the collective time evolution effects that follow after single spin energy barriers are surmounted, as illustrated in Figs. 1 and 2 of Ref. [\onlinecite{ulrich}]. Anyway, our main interest does not lie in the time dependent properties of the RAD model, but in its equilibrium behavior, which
must clearly be the same as for a system of magnetic dipoles under a dominant anisotropy with random axes. 

A summary of our results follows. We first illustrate advantages of tempered\cite{tempering0} Monte Carlo (TMC) over Metropolis \cite{metro} Monte Carlo (MMC) simulations for the calculation of equilibrium behavior. This includes a comparison of the time dependent magnetic susceptibility $\chi$ (from MMC runs),
which is characteristic of spin glasses, for the RAD model in 2D, with equilibrium results that follow from TMC simulations. We obtain equilibrium results (from TMC simulations) for systems of $L^d$ spins ($d$ is the lattice dimension) for $d=2$ and $d=3$, for $L=4,8,16$ and for $L=4,6,8,12$, respectively.
Simulations of larger systems are very time consuming, because running times grow as $L^{2d}$ for systems with {\it dipolar} interactions. 
Extrapolations to the $L\rightarrow \infty$ limit point to the following conclusions. 
In three dimensions (3D), the paramagnetic phase covers the $T>T_c$ range, where $T$ is the temperature, $T_c=( 0.86\pm0.07)\varepsilon_d/k_B$, $k_B$ is Boltzmann's constant, and $\varepsilon_d$ is a dipole-dipole nn interaction energy which is defined below, in Sec. \ref{model}. For $T<T_c$, there is an equilibrium spin glass phase. In it, the overlap parameter, as defined in Sec. \ref{olr}, is approximately given by $\sqrt{1-T/T_c}$. From our results we cannot quite 
conclude whether the droplet\cite{droplet,also} or RSB\cite{RSB,also} picture describes the RAD spin glass in 3D. Results for the correlation length $\xi$, are consistent with $\xi\sim (T-T_c)^{-\nu}$, where $T_c\simeq 0.88$ and $\nu=1.5\pm 0.5$.\cite{univ} In 2D, the paramagnetic phase covers the $T\gtrsim 0$ range, though we cannot rule out a spin glass phase below $T\simeq0.1$. Results for the correlation length $\xi$ are consistent with $\xi\sim T^{-\nu}$, where $\nu=3\pm 1$.

\section{model and method}
\label{mmth}

\subsection{Model}
\label{model}

To define the model, let
\begin{equation}
{\cal H}=\frac{1}{2}\sum_{ ij}\sum_{\alpha\beta}
T_{ij}^{\alpha\beta}S_i^\alpha S_j^\beta
\end{equation}
be its Hamiltonian, where $S_i^\alpha$ is the $\alpha$ (one of three) component of the classical spin on a cubic lattice site $i$, 
\begin{equation}
T_{ij}^{\alpha\beta}=\varepsilon_d
(a/r_{ij})^3(\delta_{\alpha\beta}-3
r_{ij}^\alpha r_{ij}^\beta/r_{ij}^2),
\label{T}
\end{equation} 
$ r_{ij}$ is the distance between $i$ and $j$, $\varepsilon_d$ is an energy, and $a$ a nn distance. Each spin points along a randomly chosen direction. More precisely,
let $\textbf{ u}_j$ be a $3-$component vector chosen randomly for each $i$ from a spherically uniform distribution of unit vectors, and let $\sigma_j=\pm 1$ at each site, such that $\textbf{ S}_j=\textbf{ u}_j\sigma_j$.
\label{sigma}
Then, ${\cal H}$ becomes,
\begin{equation}
{\cal H}=-\frac{1}{2}\sum_{ ij}J_{ij}\sigma_i\sigma_j,
\label{Hr}
\end{equation}
where
$J_{ij}=-\sum_{\alpha ,\beta}T_{ij}^{\alpha\beta}u^{\alpha}_i
u^{\beta}_j$. Thus, the RAD model is an Ising model whose bonds $J_{ij}$ are determined by the dipole-dipole terms $T_{ij}^{\alpha\beta}$ and the set of 3-component randomly oriented unit vectors $\{\bf{u}_j\}$.

We use periodic boundary conditions in 2D and 3D. Simple cubic lattices and zero applied magnetic field $H$ are assumed throughout.
We only work with $L^d$ box-like systems, and let dipole-dipole interactions act between
each spin and all other spins within an $L^d$ box centered on it. Because of the long range nature of dipolar interactions, contributions from
beyond this box would have to be taken into account (by some scheme, such as Ewald's summation) if spins were to point in any one preferred direction. 
They do not do so in this (nonferromagnetic) model as long as $H=0$. The boundary conditions as well as the $L^d$ box scheme we use here are as in Refs. [\onlinecite{nos0,nos,08}].
Finally, it is worth recalling that thermal equilibrium results obtained for H=0 for large cubic-shaped systems can, by virtue of GriffithÕs theorem \cite{griffiths} be generalized to other shapes in three dimensions.

From here on, all temperatures are given in terms of $\varepsilon_d/k_B$, where $k_B$ is Boltzmann's constant.

\subsection{Monte Carlo}
\label{tmc}

Let us first specify how we update the state of the system in all Monte Carlo evolutions. 
Initially, we compute the dipolar field at each site. Throughout a computer run, tables of all spins and dipolar fields are kept.
Dipolar fields are updated throughout all sites in the system whenever a spin is flipped. Thus, no computer time is wasted whenever an attempt to flip a spin ends in failure. This becomes important at low temperatures.

\begin{figure}[!ht]
\includegraphics*[width=80mm]{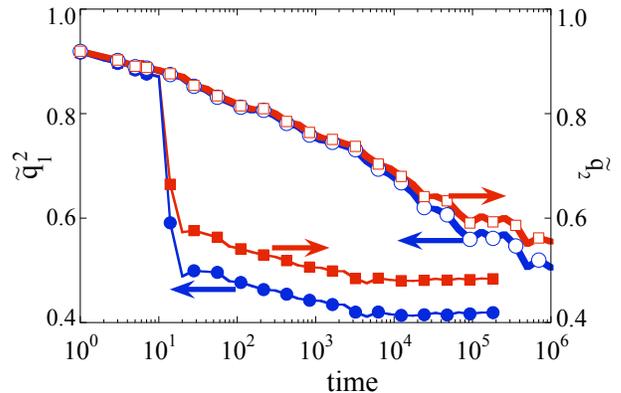}
\caption{(Color online) $\tilde {q}_1^2$ and $\tilde {q}_2$ v s time for systems of $6\times 6\times 6$ spins at $T=0.5$. $\square$ and $\circ$ are for $\tilde {q}_2$ and $\tilde {q}_1^2$, respectively. They both follow from the MMC algorithm. On the other hand, $\blacksquare$ and $\bullet$ are also for $\tilde {q}_2$ and $\tilde {q}_1^2$, respectively, but they both follow from the
TMC algorithm. All data points stand for averages over $200$ samples, each with different random anisotropy axes. All systems were allowed to evolve for over $10^5$ MCS before any measurements were taken.}
\label{vstime} 
\end{figure}

\begin{figure}[!ht]
\includegraphics*[width=80mm]{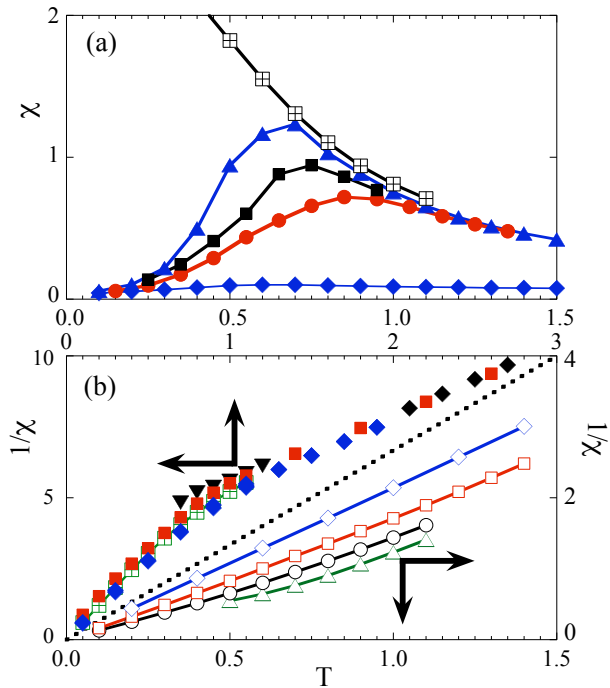}
\caption{(Color online) (a) In plane ($\chi_\|$) and out of plane ($\chi_\bot$) susceptibilities vs $T$ for 2D systems of $L\times L$ spins.  $\blacktriangle$, $\blacksquare$, and $\bullet$ are for $\chi_\|$, from MMC runs of $10^4$, $10^5$, and $10^6$ MCS, respectively. 
The low lying string of $\blacklozenge$ symbols is for $\chi_\bot$. $\boxplus$ is for data points which follow from TMC for $\chi_\|$. Finally, all $\boxplus$, $\blacktriangle$, and $\blacklozenge$ are for $L=32$ and the rest of data points are for $L=16$.   
Error bars are roughly given by icon sizes. (b) $1/\chi$ vs $T$, from TMC simulations. All data points above (below) the diagonal dotted line are for 3D (2D) systems. $\diamond$, $\square$, $\circ$, and  $\triangle$ are for $L=4, 8, 16$, and $32$, respectively. $\blacklozenge$, $\boxplus$, $\blacksquare$, and $\blacktriangledown$ are for $L=4, 6, 8$, and $12$, respectively.
Parameters for TMC runs are given in table I.} 
\label{uno} 
\end{figure}

The idea behind the tempered Monte Carlo algorithm,\cite{tempering0} is as follows. Consider two systems, 2 and 1, in thermal equilibrium, not among themselves but each one of them with its own heat bath, at temperatures $T_1$ and $T_2$, respectively. Let $T_2>T_1$, and let $E_2$ and $E_1$ be the energies of systems 1 and 2 at some given time. In the TMC algorithm, the states of two systems are exchanged with a certain probability $p$ at some specified times. It follows straightforwardly that the canonical thermal probability distributions for systems 1 and 2 are unchanged if $p=1$ if $E_2\leq E_1$, and $p=\exp [(\beta_1-\beta_2)(E_1-E_2)]$ if $E_2> E_1$, where $\beta_k=1/T_k$ for $k=1,2$.\cite{tempering0}

We do TMC simulations on $k$ identical systems at temperatures 
$T+n\Delta T$, where $n=1, 2\ldots k$, with initially independent random spin configurations, and let all systems evolve in time following the MMC algorithm for a number $\tilde{n}$ of consecutive MMC sweeps. (We choose $\tilde{n}=10$ throughout.) After every $\tilde{n}$ MMC sweeps, pairs of systems are given a chance to exchange energy, following the above given rule. More specifically, systems $2n$ and $2n+1$, for $n=0,1,2,\cdots$, are allowed to exchange states at $j\tilde{n}$ times, where $j=1,2,\cdots$, and systems $2n$ and $2n-1$, for $n=1,2,\cdots$, are allowed to exchange states at $(j+1/2)\tilde{n}$ times.
We choose $\Delta T$ as follows. Frequent exchanges take place if the energy difference $\Delta E$ between systems $2n$ and $2n\pm 1$ is not much larger than the energy fluctuations.\cite{tempering0} On the other hand, we know from our own simulations of the RAD model, that the specific heat $C$ fulfills $C\approx T^2$ for $T\lesssim 0.6$ and $C \lesssim T^2$ for all $0<T$, both in 2D and 3D. Using $C\lesssim T^2$, one obtains $\Delta T\lesssim 1/\sqrt{N}$, which is the desired condition. 

How much faster stationary states are approached in TMC than in MMC simulations is illustrated in Fig. \ref{vstime}, where plots of $\tilde{q}_1^2$ and $\tilde{q}_2$ (defined in Sec. \ref{olr}) vs time are shown, using data points from both MMC and TMC. For further comparison, results obtained for the susceptibility $\chi$ from MMC and TMC simulations are shown in Fig. \ref{uno}a. All data points, except the low lying branch, are for the ``in plane'' susceptibility $\chi_\|$, that is, the linear in plane magnetization response to an in plane applied magnetic field. The low lying branch in Fig. \ref{uno}a is for the ``out of plane'' linear susceptibility $\chi_\bot$. (As is well known, dipolar interactions lead to ``shape anisotropy'', which 
for 2D gives $\chi_\| \gg\chi_\bot$.\cite{nos}) We often write $\chi$ for $\chi_\|$. 
All data points in Figs. \ref{uno}a,  \ref{uno}b, and  \ref{uno}c follow from measurements of magnetization fluctuations in $H=0$. 

The data points from MMC simulations clearly exhibit time dependent effects that are sometimes associated with spin glasses. 
The peak in $\chi_\|$ shifts to lower values of $T$ as the number of MCS increases. This is as expected from a spin glass. Results from MMC in 3D (not shown) do not differ qualitatively from the results shown in Fig. \ref{uno}a for 2D. 
Finally, equilibrium susceptibilities that follow from magnetization fluctuations in TMC runs are shown in Fig. \ref{uno}b for system of various sizes, in 2D and in 3D.

\subsection{Overlaps}
\label{olr}

We next define the equilibrium quantities we calculate.  
Following the original idea of Edwards and Anderson,\cite{EA} consider two identical replicas, 1 and 2, of a system. Both replicas have the same set of anisotropy axes but evolve in time independently, starting from arbitrarily different initial states.\cite{EA} Let 
\begin{equation} 
\phi_j=\sigma^{(1)}_j\sigma^{(2)}_j,
\label{phi}
\end{equation}
where $\sigma^{(1)}_j$ and $\sigma^{(2)}_j$ be the spins on site $j$ of replicas 1 and 2,
and
\begin{equation} 
q= L^{-d}\sum_j\phi_j.
\label{qtilde}
\end{equation}
We also define the moments of $q$, 
$q_k= \langle \mid q \mid ^k  \rangle$,
for $k=1$, $2$ and $4$, where $\langle \ldots\rangle$ stands for an average over equilibrium states of a large number $N_s$ of replica pairs with independent random axes orientations. Note we use an absolute value in the definition of $q_1$. We refer to $q_1$ as the overlap parameter.
The spin glass susceptibility is given by $L^dq_2$.

Recall that if the probability distribution $P(q)$ in the spin glass phase differs from zero only in a vanishingly small neighborhood of some $q=\pm q_0$, where $0< q_0\leq 1$, as in the droplet model,\cite{droplet,also} then, $q_2=q_1^2>0$. On the other hand, if $P(q)\neq 0$ over a finite range of $q$ values, as in the RSB scheme,\cite{RSB,also} then $q_2>q_1^2$. 

In order to keep track of time evolutions, we also define $\tilde{\phi}_j(t_0,t)=\sigma_j(t_0)\sigma_j(t_0+t)$, in close analogy with the definition of Eq. \ref{phi}, except that both $\sigma_j(t_0)$ and $\sigma_j(t_0+t)$ are the same spin, at site $j$, at times $t_0$ and $t_0+t$, respectively. We also define $\tilde{q}(t,t_0)=L^{-d}\sum_j\tilde{\phi}_j(t_0,t)$, and the moments $\tilde{q}_k$ in obvious analogy to $q_k$. 
No measurement is ever taken, neither for the calculation of $q_k$ nor for  $\tilde{q}_k(t_0,t_0+t)$, in any simulation up to time $t_0$. The question is how to choose $t_0$. Obviously, the $t\rightarrow\infty$ limit of  $\tilde{q}_k(t,t_0)$ depends on $t_0$. Indeed, aging is the outcome of a rather long lasting dependence on $t_0$.\cite{eaging,expymcaging,mcaging} For equilibrium results, we choose sufficiently large values of $t_0$ in order that $\tilde{q}_k(t,t_0)$ reach steady state before $t=t_0$. Failure to do so would imply that equilibrium had not been reached by $t_0$, after which time measurements had been taken. We thus (a) let $t_0$ be halfway to the end of each MC run, that is, we let $t_0=t_f$, where $2t_f$ is the total number of MC sweeps taken in any given run, starting from a random spin configuration, and (b) let $t_f$ be sufficiently large for $\tilde{q}_k$ to have reached steady state by the end of the run. For short, we write $\tilde{q}_k$ for $\tilde{q}_k(t_f/2,t_f)$.
All of this is necessary but not sufficient. Conceivably, an exceedingly fast initial evolution away from a disordered state at an early $t_0$ could drive $\tilde{q}_k(t_0)$ to a null value, long before equilibrium was reached. On the other hand, the value of $q_k$, averaged over the time interval $(t_0,t_f)$, would still depend on $t_f$. Therefore, for equilibrium calculations we choose $t_0$ (and therefore $t_f$) sufficiently long for $\tilde{q}_k$ to become equal to $q_k$. For comparison, equilibrium data points for both $q_k$ and $\tilde{q}_k$ are sometimes displayed jointly.

\begin{table}
\caption{TMC simulation parameters. }
\begin{ruledtabular}
\begin{tabular}{|r|r|r|r|r|r|r|r|r|}
d, L & 2, 4 & 2, 8 & 2, 16 & 2, 32 & 3, 4 & 3, 6  & 3, 8 & 3, 12\\ 
$N_s$ & 1000 & 600 & 300 & 100  & 1800 & 800 & 400 & 175 \\
$\Delta T$ &  0.1 & 0.1 & 0.1 &  0.05 & 0.1 & 0.1 & 0.05 & 0.05 \\
MCS & $10^4$&$10^5$&$10^5$& $10^5$ &$4\times 10^4$&$2\times 10^5$&$2\times 10^5$&$2\times 10^5$\\
\end{tabular}
\end{ruledtabular}
\end{table}

\section{Equilibrium results}
We report our equilibrium results in this section. The relevant parameters for all TMC simulations from which these results follow can be found in Table I.
\begin{figure}[!ht]
\includegraphics*[width=80mm]{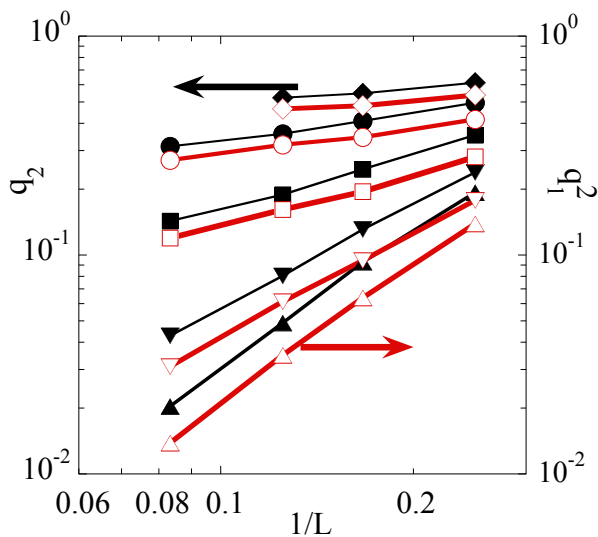}
\caption{(Color online) Log-log plots of $q_2$ (black) and $q_1^2$ (red) vs $1/L$ in 3D. Closed and open icons are for $q_2$ and $q_1^2$, respectively. $\blacklozenge$ and $\diamond$ are for $T=0.45$, $\bullet$ and $\circ$ are for $T=0.6$, $\blacksquare$ and $\square$ are for $T=0.8$, $\blacktriangledown$ and $\triangledown$ are for $T=1.0$, and $\blacktriangle$ and $\triangle$ are for $T=1.1$. Lines are guides to the eye.}
\label{3Dq2} 
\end{figure}

\begin{figure}[!ht]
\includegraphics*[width=80mm]{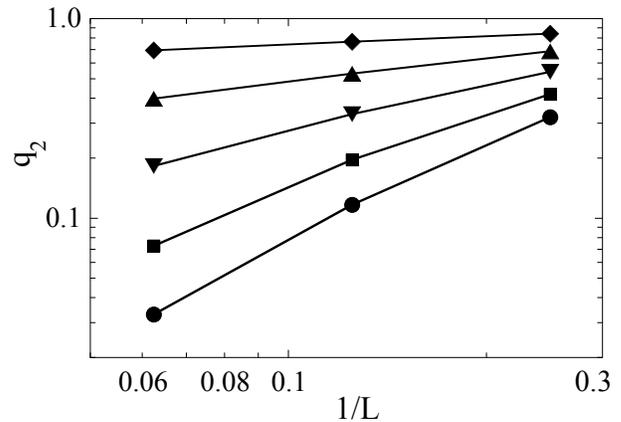}
\caption{Plots of $q_2$ vs $1/L$ in $2D$ for T=0.2 ($\blacklozenge$), 0.4 ($\blacktriangle$), 0.6 ($\blacktriangledown$), 0.8 ($\blacksquare$), 1.0 ($\bullet$). Lines are guides to the eye.}
\label{2Dq2}
\end{figure} 
Plots of equilibrium values of $q_2$ and $q_1^2$ vs $1/L$ are shown in Fig. \ref{3Dq2} for systems in 3D at various temperatures. For $T\lesssim 1$ ($T\gtrsim 1$), $q_2$ and $q_1^2$ curve upward (downward). This suggests $T_c\sim 1$. Extrapolations performed on linear plots (not shown) of $q_2$ and $q_1^2$ vs $1/L$ give $1/L\rightarrow 0$ values that are well fitted by
\begin{equation}
q_1^2=1-\frac{T}{T_c},
\label{qvsT}
\end{equation}
for $T<T_c$, and a value of $T_c$ that is well within errors of (the value we find below) $T_c=0.86\pm0.07$. In addition, 
$q_2$ and $q_1^2$ extrapolate to roughly the same value, for $T<T_c$. This would be in accordance with the droplet model of spin glasses. However, for reasons given below, this is not a firm conclusion. 

In principle, the critical exponent $\eta$ can be obtained from the plots of $q_2$ vs $1/L$ shown in Fig. \ref{3Dq2}, making use of $q_2\sim 1/L^{d-2+\eta}$ at $T_c$, which follows from finite size scaling.\cite{kawa,barber,katz} In fact, however, no meaningful number was obtained for $\eta$, because the errors turned out to be too large.

Similar plots of $q_2$ and $q_1^2$ vs $1/L$ for 2D are shown in Fig. \ref{2Dq2}. They clearly suggest that, at least for $T\gtrsim 0.4$, $q_2\rightarrow 0$ as $1/L\rightarrow 0$.

In order to examine the data we have for $q_1$ and $q_2$ in a slightly different way, we define, 
\begin{equation}
{u}_{12}=\frac{2}{\pi -2}\left(\frac{\pi} {2} -\frac{q_2}{q_1^2} \right).
\label{u12}
\end{equation}
Note that $u_{12}$ is scale free, and is consequently only a function of $\xi /L$, according to  finite size scaling (FSS) theory. \cite{kawa,barber,katz}
$u_{12}$ is analogous to Binder's ratio ${u}_{24}$, which is defined in terms of ${q}_4$ and ${q}_2$. \cite{challa} Clearly, $u_{12}=1$ for the droplet model. On the other hand $u_{12}=0$ for a macroscopic paramagnetic system, since $q$ is normally distributed then, as follows from the central limit theorem and the fact that $\xi$ is finite in a paramagnet. 

Replacement of $ q_k$ by $\tilde{q}_k$ for all $k$ in Eq. (\ref{u12}) gives the definition of $\tilde{u}_{12}$.
\begin{figure}[!ht]
\includegraphics*[width=80mm]{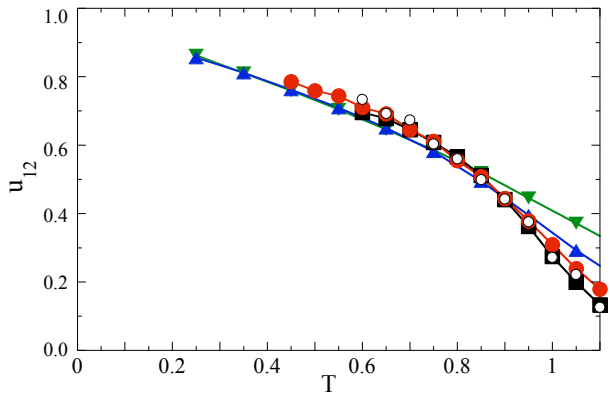}
\caption{(Color online) $u_{12}$ vs $T$ for systems of $L\times L\times L$ spins in 3D. $\blacktriangledown$, $\blacktriangle$, $\bullet$, and $\blacksquare$ are for $L=4, 6, 8$, and $12$, respectively. In addition, data points ($\circ$) for $\tilde{u}_{12}$ are given for $L=12$. Lines are guides to the eye. Error bars are roughly given by the icon sizes. } 
\label{cuatro}
\end{figure} 

\begin{figure}[!ht]
\includegraphics*[width=80mm]{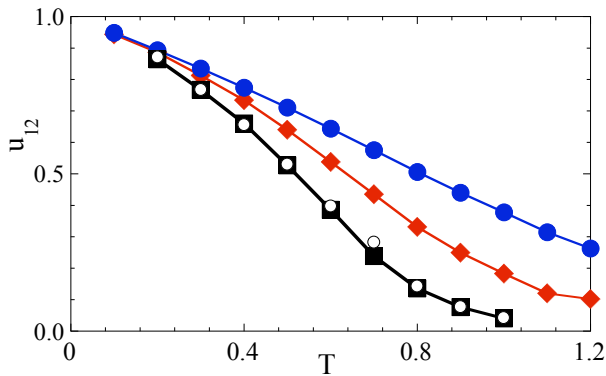}
\caption{(Color online) Plot of $ {u}_{12} $ vs $T$ for 2D systems of $L\times L$ spins, for $L=4$ ($\bullet$), $L=8$ ($\blacklozenge$), and $L=16$ ($\blacksquare$). Open icons ($\circ$) stand for $\tilde{u}_{12}$ for $L=16$. Lines are guides to the eye. Parameters for all (TMC) runs are given in table I. Icons are approximately of error bar size. }
\label{tres}
\end{figure}

From TMC simulations we obtain the equilibrium results for the RAD model in 3D that are shown in Fig. \ref{cuatro}.  Data points for $\tilde{u}_{12}$ are also shown for $L=12$ in Fig. \ref{cuatro} in order to illustrate the kind of agreement we obtain between $u_{12}$ and $\tilde{u}_{12}$. It is not clear in Fig. \ref{cuatro} whether ${u}_{12}$ becomes approximately independent of $L$ or keeps increasing with $L$ for larger values of $L$ and $ T\lesssim 0.9$. Size independence then implies $q_2/q_1^2\simeq 1+0.3T$ for $T<T_c$. On the other hand, $u_{12}\rightarrow 1$ as $L\rightarrow\infty$, would give $q_2\simeq q_1^2$ for macroscopic sizes, which would be in agreement with the tentative inference we drew from Fig. \ref{3Dq2}. We are thus led to 
\begin{equation}
1\leq \frac{q_2}{q_1^2}\lesssim 1+0.3T
\label{ineq}
\end{equation}
for macroscopic sizes, which does not discriminate between the Droplet and RSB pictures of the RAD model. 

For $T>0.9$ we have plotted (not shown) $u_{12}$ vs $1/L$, using data points from Fig. \ref{cuatro}.
Such plots point to $u_{12}\rightarrow 0$ as $1/L\rightarrow 0$, which in turn implies there is a paramagnetic phase for $T\gtrsim 0.9$. 

It is interesting to compare the above results with the ones we obtained for 2D. Plots of $u_{12}$ vs $T$ are shown in Fig. \ref{tres} for various system sizes. Data points for $\tilde {u}_{12}$ are also shown for $L=16$. In contrast with the results for 3D, the three curves in Fig. \ref{tres} appear to come together only gradually, as $T\rightarrow 0$. Plots (not shown) of $u_{12}$ vs $1/L$, can be made from the the data points shown in Fig.  \ref{tres}. One can then extrapolate  $u_{12}$ to $1/L\rightarrow 0$. At least for $0.2\lesssim T$, $u_{12}\rightarrow 0$, which is consistent with a paramagnetic phase.

\begin{figure}[!ht]
\includegraphics*[width=80mm]{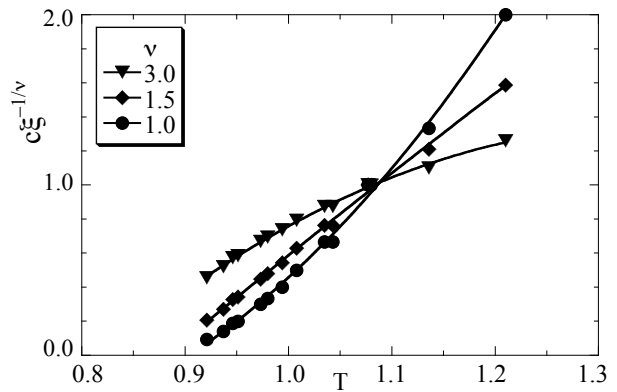}
\caption{Plots of $c\xi^{-1/\nu}$ vs $T$ for $3D$ and the values of $\nu$ that are given in the graph. $c$ is some undetermined constant. Lines are guides to the eye.}
\label{inset3}
\end{figure}

\begin{figure}[!ht]
\includegraphics*[width=80mm]{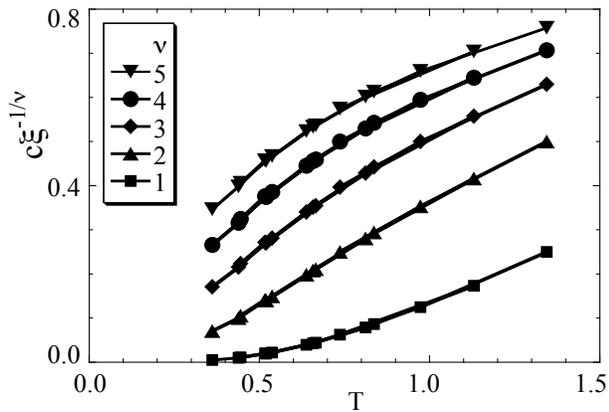}
\caption{Plots of $c\xi^{-1/\nu}$ vs $T$ for $2D$ the values of $\nu$ that are given in the graph. $c$ is some undetermined constant. Lines are guides to the eye.}
\label{2DXi}
\end{figure}

We can obtain $\xi$ (of an infinite size system) making use of the data for $u_{12}$ and of the fact that, according to FSS,\cite{barber,katz}
$u_{12}$ is only a function of $\xi  /L$. Note that $\xi/L$ is constant for any horizontal line that intersects the all the curves in either Fig. \ref{cuatro} or Fig. \ref{tres}. We can thus obtain, $\xi (T_n)/L_n=c$, where $c$ is some constant and $T_n$ and $L_n$ are the values of $T$ and $L$ where a horizontal line crosses the $nth$ curve in Figs. \ref{cuatro} or \ref{tres}. Different horizontal lines give different values of $c$ which can be chosen independently in order to collapse all 
plots of $\xi$ vs $T$ into a single $\xi$ vs $T$ curve. Thus, we obtain the plots shown in Figs. \ref{inset3} and \ref{2DXi}.

Extrapolations in plots such as the ones shown in Fig.\ref{inset3} give $T_c\simeq 0.88$ for 3D. From the errors in the data for $u_{12}$, we estimate an error $\delta T_c=0.05$. We determine the exponent $\nu$, in $\xi \sim (T-T_c)^{-\nu}$, from these plots. The value $\nu\simeq 1.5$ gives the best straight line fit in the $0.88<T<1.2$ range. On this basis we adopt the value $\nu\simeq 1.5$. Fits obtained from $\nu$ values outside the $1\lesssim \nu\lesssim 2$ range show significant curvature, whence we assign the error $\delta\nu=0.5$. Proceeding similarly for 2D, using plots as the ones shown in Fig. \ref{2DXi}, we obtain $T_c\simeq 0$, though a spin glass phase below $T\simeq0.1$ is conceivable, and $\nu=3\pm 1$. 

From different extrapolation procedures we have arrived at values of $T_c$ in the [0.83,0.88] range. Considering all the errors involved, we arrive at
\begin{equation}
T_c=0.86\pm 0.07
\label{Tc}
\end{equation}
for the RAD model in 3D.

\section{conclusions}

In sum, we have studied the equilibrium behavior of the RAD model by means of tempered Monte Carlo simulations. The sizes of the systems we have simulated, temperatures, as well as other parameters, are given in Table I. From them, we have drawn quantitative evidence that points to the following conclusions.
In 3D, the paramagnetic phase covers the $T>T_c$ range, where $T_c= 0.86\pm0.07$. For $T< T_c$, there is a spin glass phase. In it, the overlap parameter, defined in Sec. \ref{olr}, does not vanish. It is approximately given by Eq. (\ref{qvsT}). No information about critical behavior should be drawn from this equation, because it is not sufficiently accurate for it.
From extrapolations of $q_2$ and $q_1^2$ to the $1/L\rightarrow 0$ limit (see Fig. \ref{3Dq2}), one might be tempted to infer that $q_2$ and $q_1^2$ become then equal, as in the droplet model. 
However, plots of $u_{12}$ vs $T$, shown in Fig. \ref{cuatro}, do not provide firm support for such a conclusion, because the $L\rightarrow\infty$ limit of $u_{12}$ in the spin glass phase seems uncertain. We can only be reasonably sure that the limit is somewhere between the value of $u_{12}$ shown for $L=12$ and $1$. From this, Eq. (\ref{ineq}), which does not discriminate between the applicability of the droplet\cite{droplet} or RSB\cite{RSB} pictures to the RAD model, follows.  
Results for the correlation length $\xi$, exhibited in Fig. \ref{inset3}, are consistent with $\xi\sim (T-T_c)^{-\nu}$, where $\nu=1.5\pm 0.5$.\cite{univ}

In 2D, the paramagnetic phase covers the $T\gtrsim 0$ range, though we cannot rule out a spin glass phase below $T\simeq0.1$. Results for $\xi$, exhibited in Fig. \ref{2DXi}, are consistent with $\xi\sim T^{-\nu}$, where $\nu=3\pm 1$.

\acknowledgments
We had interesting discussions with J. J. Alonso. We are indebted to the BIFI Institute (at Universidad de Zaragoza) and to the Carlos I Institute (at Universidad de Granada)
for letting us run on many of their cluster nodes for months, and for financial support from Grant No. FIS2006-00708, from the Ministerio de Educaci\'on y Ciencia of Spain.


\begin{thebibliography}{99}
\bibitem{univ}This value of $\nu$ is within the range of the long list of values that have been obtained for short range Ising models with various random bond distributions.\cite{katz} Screening of the (long ranged) dipolar interactions might bring this about.
\end{thebibliography}
\end{document}